\documentclass{iitpressproc}

\usepackage{graphicx}
\usepackage{amsmath}
\usepackage{amssymb}

\begin{document}

\title{Diffraction at the LHC}

\author{M. Csan\'ad
\address{E\"otv\"os University, Budapest, Hungary}
}

\maketitle

\begin{abstract}
Proton-proton collisions at the LHC can be classified as elastic, non-diffractive, and diffractive.
In this paper we discuss various measurements of these above processes at various
LHC experiments. We report about the total proton-proton cross-section measurements, about
the analysis of diffractive events and also about the pseudorapidity distribution in inelastic events.
\end{abstract}

\section{Introduction}
High energy proton-proton collisions are observed at the Large Hadron Collider of CERN.
A large fraction of these collisions are events where color degrees of freedom are exchanged
between the two protons, and new particles are created throughout the phase-space spanned by
longitudinal pseudorapidity, defined as $\eta=0.5\ln[(p+p_z)/(p-p_z)]$ and azimuth angle defined as
$\phi = \arctan(p_y/p_x)$.
However, a substantial fraction of the total pp cross section is due to diffractive processes,
where protons remain intact or dissociate into a mixture of particles with a low diffractive mass (compared to
the original energy of the proton). These events can be characterized by a colorless momentum
exchange of the two protons, via the Pomeron, carrying the quantum numbers of the vacuum.
A significant number of elastic events are also observed at the LHC, where Pomeron exchange governs
the deflection of the protons, both of which remain intact, and no new particles are created (i.e.
the kinetic energy is retained in the collision).

\begin{figure}
  \begin{center}
  \includegraphics[width=1.0\linewidth]{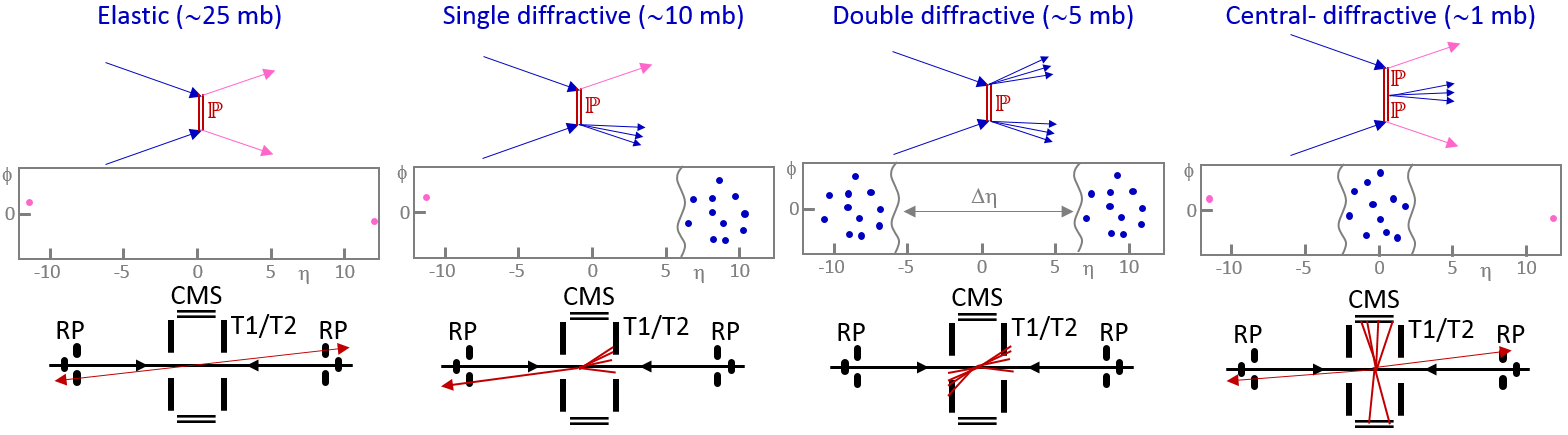}
\caption{Event classification in the TOTEM+CMS detector system. In elastic pp events, only momentum
is exchanged (by colorless particles), while in diffractive events, particles are produced either through
the dissociation of one or both of the protons, or through the interaction of the exchanged Pomeron.
\label{f:events} }
  \end{center}
\end{figure}

A summary of diffractive p+p event classes is shown in Fig.~\ref{f:events}. The particles in the final state
are separated by a large ($\gtrsim 3$) rapidity gap $\Delta \eta$\footnote{E.g. in single
diffraction, a rapidity gap means no particles in a pseudorapidity region defined by $\Delta \eta$
between the leading proton and the diffractive system.}
The rapidity gap size is related to the diffractive mass of the dissociated
system, $M_X$. In case of single diffraction, $\Delta \eta \approx  \ln \xi$, with $\xi = \Delta p/p$,
being the momentum loss of the proton, also equal $\xi=M_X^2/s$ where $s=(p_1+p_2)^2$, i.e. the
total center of mass energy squared. In order to experimentally classify these events, one needs to 
identify the rapidity gap in particle production, with detectors covering a large rapidity region.

The LHC TOTEM experiment detects the leading protons with a system
of Roman Pots (RP). There are four RP stations, $\pm$147 and $\pm$220 meters from the Interaction Point (IP5),
with two units at each location. A unit consists of two vertical and one horizontal edgeless Si strip detector~\cite{Ruggiero:2009zz},
reaching to a distance of a few millimeters from the beam.  RP's allow
TOTEM to track protons which were deflected by a few microradians only (in elastic, single diffractive
or central diffractive events). Two tracking detectors at a smaller pseudorapidity
allow TOTEM to track almost the entirety of diffractive dissociation. These are 
the T1 and T2 telescopes, covering $3.1<|\eta|<4.7$ and $5.3<|\eta|<6.5$ respectively.

Located also at IP5, CMS also has a set of forward calorimeters which can be used in
observing single, double or central diffractive events. The Hadron Forward
calorimeter (HF) is at 3$<|\eta|<$5, while the CASTOR calorimeter
is located only on one side of the IP, at $-6.6<|\eta|<-5.2$.

At ATLAS, there are forward calorimeters outside the tracking system. A highly segmented electromagnetic
(EM) liquid argon sampling calorimeter covers the range $|\eta|< 3.2$.
The Hadronic End-cap Calorimeter  (HEC) covers $1.5 < |\eta| < 3.2$,
while the Forward Calorimeter (FCal) covers $3.1 < |\eta|< 4.9$. For triggering,
ATLAS uses Minimum Bias Trigger Scintillators (MBTS, $2.1 < |\eta| < 3.8$), placed
on the inner face of the HEC. For the analysis of rapidity gaps, only calorimeter information is used
in the region $2.5 < |\eta| < 4.9$, beyond the acceptance of the inner detector.

\section{Cross-section measurements}
\begin{figure}
  \begin{center}
  \includegraphics[width=1.0\linewidth]{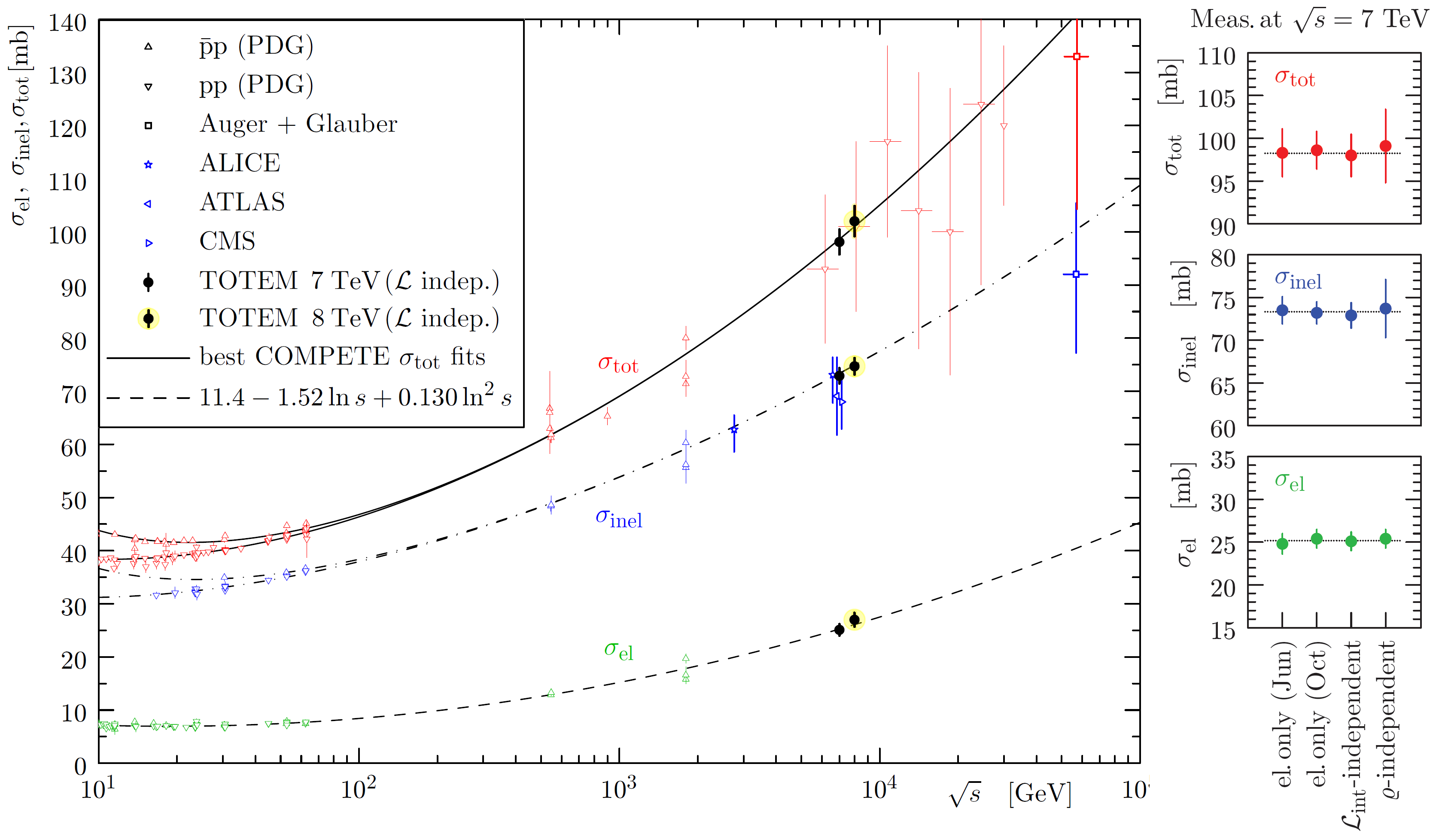}
\caption{TOTEM luminosity-independent p+p cross-section measurements with
world cross-section data as a function of $\sqrt{s}$. On the right, comparison of the TOTEM measurements
obtained with different analysis methods are shown.\label{f:sigmatot} }
  \end{center}
\end{figure}

Cross-section of pp collisions is a particularly interesting, since the
total cross-section rises with increasing total center of mass energy ($\sqrt{s}$),
which can only be explained in terms of the Regge theory with the exchange of Pomerons.
Measurements of $\sigma_{\rm tot}$ at LHC energies were available only in high uncertainty
cosmic-ray data. TOTEM has now measured the total, elastic and inelastic pp cross-sections
 with a very high precision, with several independent methods (see Fig.~\ref{f:sigmatot} and Refs.~\cite{Antchev:2011vs,Antchev:2013gaa,Antchev:2013iaa,Antchev:2013haa,Antchev:2013paa}):\\
a) The basic method is to measure the
number of elastic and inelastic events ($N_{\rm el, inel}$), and using luminosity $\mathcal{L}$\footnote{Luminosity
is the number of interactions per effective area and per second}:
$\sigma_{\rm tot} = (N_{\rm el} + N_{\rm inel})/\mathcal{L}$.\\
b) There is also a method which does
not require the knowledge of the number of inelastic events, but is based instead of the optical theorem:
$\sigma_{\rm tot}^2 = 16\pi(\hbar c)^2/(1+\rho^2) \times \mathcal{L}^{-1} \times \left.dN_{\rm el}/dt\right|_{t=0}$,
where $-t$ is the momentum exchange squared, while $\rho$ is the real over imaginary part of the forward
scattering amplitude. \\
c) The third, luminosity independent method can be summarized as
$\sigma_{\rm tot} = 16\pi(\hbar c)^2/(1+\rho^2) \times (\left.dN_{\rm el}/dt\right|_{t=0})/(N_{\rm el} + N_{\rm inel})$.

As described in the previous paragraph, for the total cross-section measurement, differential
measurements of $d\sigma/dt$ are needed at small $t$ values. The minimum reachable $t$ by TOTEM
depends on the accelerator optics (described by $\beta^*$, the betatron amplitude taken at the interaction
point). If $\beta^*=3.5$ m, the range of $|t|>0.36$ GeV$^2$ can be measured, while with a special, $\beta^*=90$ m optics, ranges
of $|t|>0.005$ GeV$^2$ can be seen. With these measurements~\cite{Antchev:2013gaa,Antchev:2011zz}, extrapolation
to $t\rightarrow 0$ is possible. TOTEM has performed also a measurement at $\beta^*=1000$ m, which allowed to reach
$|t|=6\cdot10^{-4}$ GeV$^2$ where the Coulomb-nuclear interference can be studied~\cite{Kaspar:2013tda}.


In ATLAS, inelastic cross-sections are measured as a function of the accessible $\xi$ range~\cite{Aad:2012pw}.
Determined by the $\eta$ acceptance, a $\xi>5\cdot 10^{-6}$ range is accessible.
Event selection is based on triggering via the MBTS. This way a measurement with the
integrated luminosity of 20.3 $\mu$b was performed, and an inelastic cross-section
of $60.3\pm0.5$(syst)$\pm2.1$(lumi) mb was measured~\cite{Aad:2011eu}, for $\xi>5\cdot 10^{-6}$, corresponding to the
diffractive mass being larger than 15.7 GeV. A dependence on $\xi_{\rm cut}$ has also
been analyzed, from which (via extrapolation) $\sigma_{\rm inel}=69.4\pm2.4$(exp)$\pm6.9$(extr) mb was extracted~\cite{Aad:2012pw}.
Measurement of total inelastic cross-sections was also performed at ALICE~\cite{Abelev:2012sea}. The
2.76 TeV result is $62.8^{+2.4}_{-4.0}{\rm (model)}\pm1.2{\rm (lumi)}$, while the
7 TeV result, as shown in Fig.~\ref{f:sigmatot}, is $73.2^{+2.0}_{-4.6}{\rm (model)}\pm2.6{\rm (lumi)}$.
For $\xi>5\cdot 10^{-6}$, ALICE obtained $62.1^{+1.0}_{-0.9}{\rm (model)}\pm2.2{\rm (lumi)}$. The results
of all LHC experiments are consistent with each other.

\section{Single diffraction}

TOTEM measures single diffractive events with a large rapidity gap between the outgoing proton and
the dissociative hadronic system~\cite{Oljemark:2013wsa}. The final goal of this preliminary measurement
is to determine the integrated cross-section and the differential $d\sigma/dt$ distribution in each mass range. 
Event selection is done via the RP's and the Telescopes, requiring one proton and a rapidity gap in T2 on the side
of the proton, and tracks in the opposite side. The correspondence of the gap size to the diffractive mass is given in
Table~\ref{t:singlediff}. Raw rates have to be corrected for efficiency and acceptance, and backgrounds have to be subtracted.
The main background source is the pileup of a proton in RP from the beam halo with an independent non-diffractive event in T2.
Preliminary results in the $3.4 - 1100$ GeV diffractive mass region indicate $\sigma_{SD} = 6.5\pm1.3$ mb for both arms, while
differential cross-section exponential slopes are estimated to be $B=10.1, 8.5$ and $6.8$ GeV$^{-2}$
for the first three mass ranges, respectively.

\begin{table}
\begin{center}
\caption{Single diffraction event classes in TOTEM \label{t:singlediff}}
\begin{tabular}{|l|l|c|c|}
\hline
Event class  & Event topology & Mass range & Mom. loss ($\xi$) \\
\hline
Low mass & RP, opposite T2 & 3.4 -- 8 GeV & $2\cdot 10^{-7}$ -- $10^{-6}$ \\
Medium mass & RP, opp. T2, opp. T1 & 8 -- 350 GeV & $10^{-6}$ -- 0.25\%\\
High mass & RP, opp. T2, same T1 & 0.35 -- 1.1 TeV & 0.25\% -- 2.5\%\\
Very high mass & RP, opp. T2, same T2 & 1.1 TeV -- &2.5\% --\\
\hline
\end{tabular}
\end{center}
\end{table}

\begin{figure}
  \begin{center}
  \includegraphics[width=0.67\linewidth]{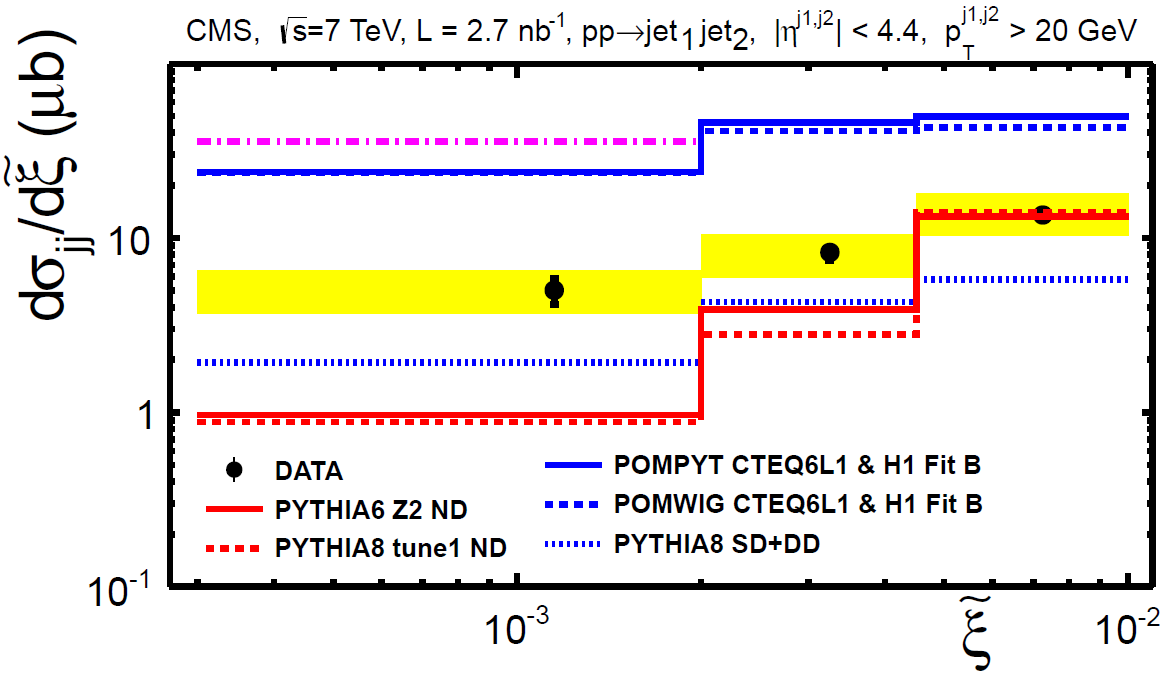}
\caption{The differential cross section for inclusive dijet production from Ref.~\cite{Chatrchyan:2012vc}.
The points are plotted at the center of the bins. The error bars indicate the statistical uncertainty and the band
represents the systematic uncertainties added in quadrature. See text and Ref.~\cite{Chatrchyan:2012vc} for details.\label{f:dijet} }
  \end{center}
\end{figure}

CMS has measured single diffractive dijet production~\cite{Chatrchyan:2012vc},
where one of the protons remains intact or is excited into a low-mass state, while
the other participates in a hard scattering with a Pomeron from the non-dissociating proton.
CMS measures the cross section for dijet production as a function of an evaluation of $\xi$ (CMS has
no direct access to the momentum loss of the undissociated proton), based on events  with at
least two jets with a transverse momentum of $p_t^{j1,j2}>20$ GeV  in the pseudorapidity
region $|\eta^{j1,j2}|<4.4$. The results are compared to diffractive and non-diffractive MC models (see Fig.~\ref{f:dijet}).
The low-$\xi$ data show a significant contribution from diffractive dijet production, observed for the first time at the LHC.
POMPYT and POMWIG, based on dPDFs from HERA, overestimate the measured cross section by a factor of 5. This
factor can be interpreted as the effect of the rapidity gap survival probability.

\section{Double and central diffraction}
TOTEM has measured the double diffractive cross-section in the very forward region at 7 TeV. With the T1 and T2 telescopes
a clean sample of double diffractive events could be extracted. The event topology was defined as no tracks
in the T1 detectors (to ensure that no particles are present at low $\eta$), and valid tracks in both T2 arms,
which selected events that have two diffractive systems
having particles with a minimal pseudorapidity in the  $4.7<|\eta|<6.5$ range. These events make up only
roughly 3\% of the total double diffractive cross-section, they however provide a pure selection of such events.
The cross-section in the total region was estimated to be $\sigma_{\rm DD}=({116 \pm 25})$ $\mu$b~\cite{Antchev:2013any}.
To access further experimental details, the above $\eta$ range was divided into two sub-regions on each
side, by cutting at $|\eta|=5.9$. In each measurement category, the selected sample had to be corrected
for trigger-efficiency, pile-up and T1 multiplicity, and also the background had to be estimated. The visible
cross-section was then determined by correcting the raw rate for acceptance and detection efficiency. Lastly,
the visible cross-section had to be corrected so that both diffractive systems have a minimal pseudorapidity
in the $4.7<|\eta|<6.5$ range. See results in Table~\ref{t:doublediff}. In order to determine the background, three event classes were considered:
non-diffractive (ND), single diffractive (SD) and central diffractive (CD). A data-driven background estimation
was used, where the values of the background estimates were calculated iteratively (see details in Ref.~\cite{Antchev:2013any}).

\begin{table}
\begin{center}
\caption{Double diffractive cross-section measurements in the forward region. Both visible and true 
$\eta_{min}$ corrected cross-sections are given. There were two subregions defined;
(1): $4.7<|\eta|_{\rm min}<5.9$  and (2): $5.9<|\eta|_{\rm min}<6.5$. D$_{ij}$ requires
then $|\eta|_{\rm min}$ to be in region $i$ on the $+$ side and in region $j$ on the $-$ side.
\label{t:doublediff}}
\begin{tabular}{|l|c|c|c|c|c|}
\hline
$\sigma_{DD}$ [$\mu$b] & Total &D$_{11}$ & D$_{22}$ & D$_{12}$ & D$_{21}$ \\
\hline
Visible & 131$\pm$22 & 58$\pm$14 & 20$\pm$8 & 31$\pm$5 & 34$\pm$5 \\
\hline
True $\eta_{min}$ & 116$\pm$25 & 65$\pm$20 & 12$\pm$5 & 26$\pm$5 & 27$\pm$5  \\
\hline
Pythia true $\eta_{min}$  & 159 & 70 & 17 & 36 & 36 \\
Phojet true $\eta_{min}$  & 101 & 44 & 12 & 23 & 23 \\
\hline
\end{tabular}
\end{center}
\end{table}

\begin{figure}
  \begin{center}
  \includegraphics[width=1.0\linewidth]{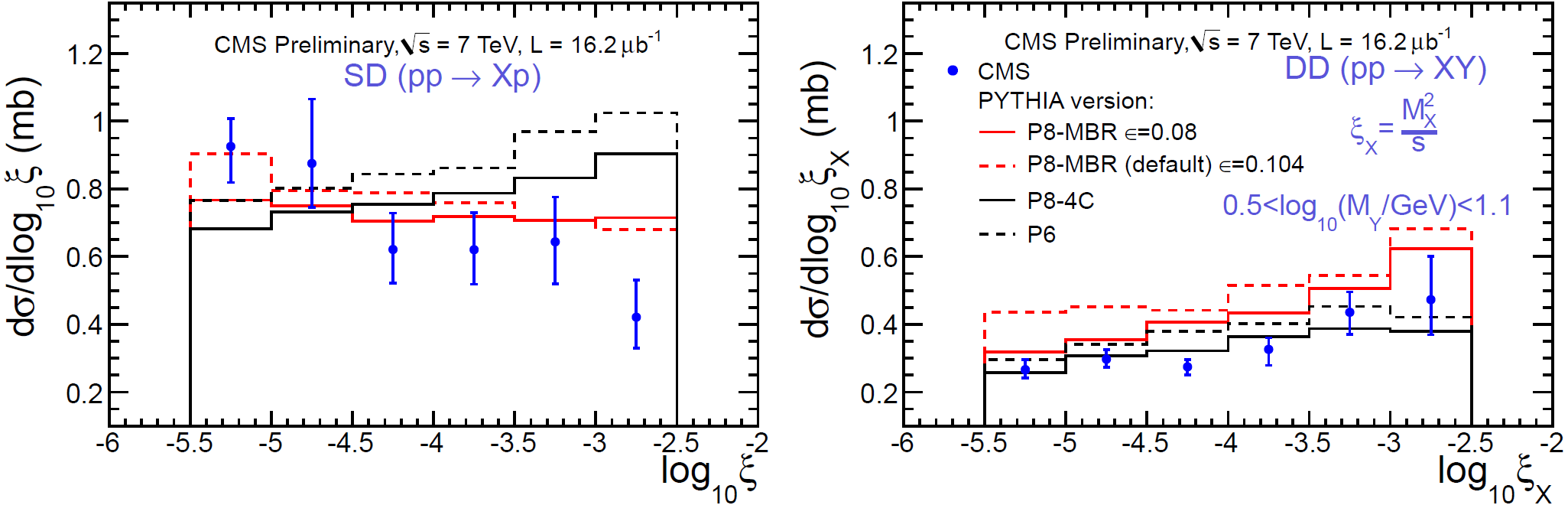}
\caption{The SD (left) and DD (right) cross sections as a function of $\xi$, compared to several MC 
predictions. \label{f:CMSdiffr} }
  \end{center}
\end{figure}

CMS has measured single and double diffractive events with the experimental topologies of a large rapidity
gap~\cite{CMS:2013mda}. Event types were defined as SD1 (gap on $+$ side), SD2 (gap on $-$ side)
and DD (central gap). 
Both SD1 and SD2 event topologies (defined as $\eta_{\rm min}>-1$ and $\eta_{\rm max}<1$, respectively) 
contain SD and DD events, with the second hadronic system being outside the central region.
For the SD2 topology, CASTOR  was used for tagging to select a DD enhanced event sample and to calculate
SD and DD cross-section, using a $\Delta\eta^0>3$ selection.  Since there is no similar detector on the
$+$ side, the SD1 sample was treated as a control sample.
The differential SD cross-section was then measured as a function of $\xi$, in the $-5.5<\log_{10}\xi<-2.5$ range,
while the DD cross-section was measured in events in which a hadronic system is detected in the central
arm ($12<M_X<394$ GeV) and the other in CASTOR ($3.2<M_Y<12$ GeV), in the $\xi_X=M_X^2/s$ range
of $-5.5<\log_{10}\xi<-2.5$. The DD cross-section was also measured differentially as a function of the
central pseudorapidity gap $\Delta\eta$. Results were compared to predictions of theoretical models,
as shown in Fig.~\ref{f:CMSdiffr} (see details in Ref.~\cite{CMS:2013mda}). In total, a
$\sigma_{SD}^{\rm vis}= 4.27 \pm 0.04$(stat.)$^{+0.65}_{-0.58}$(syst.) mb
 integrated cross-section was measured (multiplied by two to account for both side processes).
Note also, that for all the above event types, the forward rapidity gap cross-section $d\sigma/d\Delta\eta^F$
was measured both at CMS and at ATLAS. At small  $\Delta\eta^F$ , the cross-sections are dominated by 
non-diffractive events, exponentially suppressed as  $\Delta\eta^F$ increases. For $\Delta \eta^F>3$,
cross-section per unit rapidity gap weakly changes with $\Delta \eta^F$, taking the value of
$d\sigma/\Delta\eta^F \approx 1.0$ mb~\cite{Aad:2012pw}. The large $\eta$ range comparison between data and MCs allows to tune
the relative fraction of the different components in the models. CMS extends the ATLAS measurement by
0.4 unit of rapidity gap size, and the two measurements agree within uncertainties~\cite{Aad:2012pw,CMS:2013mda}.

Central diffractive events are also measured at TOTEM, defined by the event topology of
protons in both RP arms and tracks in the T2 telescopes. 
The background of this selection is mainly elastic events and inelastic events with pile-up of beam-halo.
Since during data taking the RP distance to the beam was greater than $11\sigma_{\rm beam}$,
it was found that beam halo contribution is negligible. Elastic event events can rejected
via anti-elastic cuts in the RP's, or by selecting non-elastic topologies (e.g. protons
being in the top RP's on both sides). 
Single-arm differential event rate was then measured and
a preliminary exponential slope parameter of $B=-7.8\pm1.4$ GeV$^2$ was extracted via a MC
fit on the $t_y$ distribution. The preliminary cross-section estimate of TOTEM is $\sigma_{CD} \approx 1$ mb.
CMS and TOTEM are also working on a common central diffraction measurement, where the diffractive central
mass can be measured by both experiments: using the proton tracks by TOTEM (via $M_X^2 = \xi_1\xi_2 s$) and
directly by CMS. This will provide a result with an unprecedented rapidity coverage. 

\section{Inelastic pseudorapidity distributions}

\begin{figure}
  \begin{center}
  \includegraphics[width=1.0\linewidth]{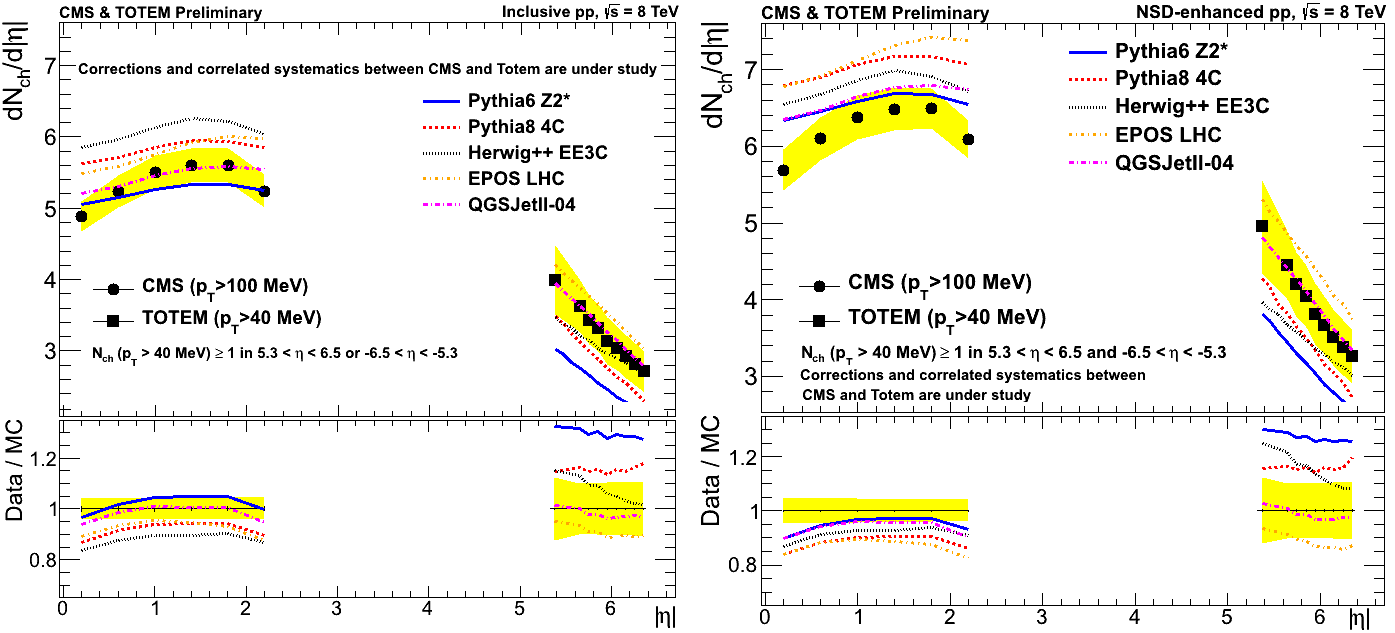}
\caption{Charged particle pseudorapidity density distribution as measured by the CMS tracker in the central region
and TOTEM T2 in the forward region for the same events with one charged particle with $p_t>40$ MeV$/c$,
for both the inclusive and the non-single diffractive enhanced sample, with comparisons to various MC
tunes (see details in~\cite{TOTEM-NOTE-2013-001}).
\label{f:TOTEMCMS_dndeta} }
  \end{center}
\end{figure}

A measurement of the pseudorapidity density (dN$_{ch}$/d$\eta$) of forward charged particles
in inelastic events was also performed by TOTEM with the T2 detectors~\cite{Aspell:2012ux}.
The measurement refers to events with at least one  charged particle with $p_t \ge 40$ MeV$/c$ and
with a mean lifetime $\tau> 0.3\times 10^{-10}$ s, directly produced in pp interactions or in subsequent
decays of particles having a shorter lifetime. Results of this measurement were compared to MC expectations:
none of the theoretical models has been found to fully  describe the data.
The cosmic ray MC generators (e.g. QGSJETII, see a review of event
generators in Ref.~\cite{d'Enterria:2011kw}) show a better agreement for the slope.
The TOTEM experiment has also measured the pseudorapidity density of charged particles with 
$p_t \ge 40$ MeV$/c$ in pp collisions at 8 TeV for $5.3<|\eta|<6.4$ in a low
intensity run with common data taking with the CMS experiment~\cite{TOTEM-NOTE-2013-001,CMS:2013yca}.
This represents the extension of the analogous measurement described in the beginning of this paragraph.
The analysis was performed on 3 different events categories: an inclusive one with at least one charged particle in either
$-6.5<\eta<5.3$ or $5.3 <\eta < 6.5$, a non-single diffractive enhanced one with at least one charged
particle in both $-6.5<\eta<5.3$ and $5.3 <\eta < 6.5$ and a single diffractive enhanced one with at
least one charged particle in $-6.5<\eta<5.3$ and none in $5.3 <\eta < 6.5$ or vice-versa.
A minimum bias trigger was provided by the TOTEM T2 telescope and contributed to the CMS Global Trigger decision
to  initiate the simultaneous read out of both CMS and TOTEM detectors. The event data from TOTEM and CMS were
then combined offline in a common data analysis activity.
The visible cross section for events triggered by T2 has been estimated to be about 95\% of the total
inelastic cross section: more than 99\% of all non-diffractive events and all single and double diffractive
events having at least one diffractive mass larger than $\approx$ 3.6 GeV$/c^2$. See results in
 Fig.~\ref{f:TOTEMCMS_dndeta}, and details about the compared event generators in 
 Refs.~\cite{TOTEM-NOTE-2013-001,CMS:2013yca}.

\section*{Acknowledgments}
M. Cs. would like to thank the Organizers of ISMD 2013 for organizing an inspiring conference. The author
is also thankful for Marco Bozzo and the TOTEM Editorial Board for their help in improving the
manuscript. The work presented in this paper was done by the TOTEM, CMS and ATLAS collaborations.
The participation of the author at ISMD was supported by the Hungarian OTKA grant NK 101438.

\bibliographystyle{prlstyl}
\bibliography{../../../Master}

\end{document}